\documentclass[twocolumn,aps,prr,showpacs,raggedbottom,nobalancelastpage,amssymb,superscriptaddress]{revtex4-1}
\usepackage[utf8]{inputenc}
\pdfoutput=1
\usepackage{color}
\usepackage{graphicx}
\usepackage{physics}
\usepackage{braket}
\usepackage{mathtools}
\usepackage[utf8]{inputenc}
\usepackage{siunitx}

\def\Tr{\mathrm{Tr}}

\newcommand{\epscj}{\epsilon_\text{c}^{(j)}}
\newcommand{\td}{t_{\text{d}}^{\phantom\dag}}
\newcommand{\tdn}{t_{\text{d}n}^{\phantom\dag}}
\newcommand{\tdL}{t_{\text{L}}^{\phantom\dag}}
\newcommand{\tcB}{t_{\text{B}}^{\phantom\dag}}
\newcommand{\tdR}{t_{\text{R}}^{\phantom\dag}}
\newcommand{\tdl}{t_{\ell}}

\newcommand{\tcn}{t_{\text{c}n}^{\phantom\dag}}

\newcommand{\Hdot}{H_{\text{dot}}^{\phantom\dag}}

\newcommand{\cdag}{c^{\dag}}

\renewcommand{\ddag}{d^{\dag}}
\newcommand{\fdag}{f^{\dag}}
\newcommand{\cpdag}{c^{\phantom\dag}}

\newcommand{\fpdag}{f^{\phantom\dag}}

\newcommand{\vecsig}{\boldsymbol{\sigma}}

\begin{document}
\title{Entanglement-based observables for quantum impurities}

\author{Lidia~Stocker}
\affiliation{Institute for Theoretical Physics, ETH Zurich, 
	8093 Zurich, Switzerland}
\author{Stefan~H.~Sack}
\affiliation{Institute for Theoretical Physics, ETH Zurich, 
	8093 Zurich, Switzerland}
\affiliation{IST Austria, Am Campus 1, 3400 Klosterneuburg, Austria}
\author{Michael~S.~Ferguson}
\affiliation{Institute for Theoretical Physics, ETH Zurich, 
	8093 Zurich, Switzerland}
\author{Oded~Zilberberg}
\affiliation{Institute for Theoretical Physics, ETH Zurich, 
	8093 Zurich, Switzerland}
\affiliation{Department of Physics, University of Konstanz, D-78457 Konstanz, Germany}

\begin{abstract}
	Quantum impurities exhibit fascinating many-body phenomena when the small interacting impurity changes the physics of a large noninteracting environment. The characterisation of such strongly correlated non-perturbative effects is particularly challenging due to the infinite size of the environment, and the inability of local correlators to capture the build-up of long-ranged entanglement in the system. Here, we harness an entanglement-based observable – the purity of the impurity – as a witness for the formation of strong correlations. We showcase the utility of our scheme by exactly solving the open Kondo box model in the small box limit, and thus describe all-electronic dot–cavity devices. Specifically, we conclusively characterise the metal-to-insulator phase transition in the system and identify how the (conducting) dot-lead Kondo singlet is quenched by an (insulating) intra-impurity singlet formation. 
	Furthermore, we propose a experimentally feasible tomography protocol for the  measurement of the purity, which motivates the observation of impurity physics through their entanglement build up. 
\end{abstract}
\maketitle

\section{Introduction}
Quantum many-body systems are an important frontier of physics~\cite{bruus_many-body_2004, coleman_introduction_2015}. The system's ground state can be dominated by many-body interactions and thus profoundly removed from its simpler single-particle limit, i.e., nonlocal entanglement manifests between the system's constituents. Correspondingly, finding the strongly correlated ground state is equivalent to solving complex optimisation problems~\cite{hauke_perspectives_2020}. Alongside theoretical and numerical studies of quantum many-body systems, analogue quantum simulators serve as an alternative method to emulate strongly correlated effects using experimentally controllable devices~\cite{georgescu_quantum_2014,preskill_quantum_2018,daley_practical_2022}. Examples of such simulators include trapped atoms~\cite{bloch_quantum_2012} or ions~\cite{dehmelt_experiments_1990,paul_electromagnetic_1990}, light-matter systems~\cite{haroche_nobel_2013}, and electronic~\cite{ihn_electronic_2004} or superconducting~\cite{houck_-chip_2012} mesoscopic devices. These comprise the bulk of the activity in the so-called noisy intermediate-scale quantum (NISQ) era~\cite{preskill_quantum_2018}. Interestingly, within the NISQ domain, entanglement has become a standard order parameter for the study of many-body physics~\cite{mandel_controlled_2003,besse_realizing_2020,greiner_quantum_2002,pichler_thermal_2013}.

Among strongly correlated systems, quantum impurity problems hold a prominent position~\cite{ashida_quantum_2020}. They engender a simple case where many-body interactions manifest only at a specific spot in space, i.e., at the quantum impurity. Nevertheless, by coupling the impurity to a large reservoir of noninteracting particles, strong correlations manifest in the whole system. As such, understanding quantum impurity problems is key to understanding quantum many-body systems as a whole~\cite{kotliar_strongly_2004,gull_continuous-time_2011}. Moreover, quantum impurities also model a large class of quantum simulators, which are composed of confined regions of space embedded within a larger experimental environment. Conversely, NISQ devices can be harnessed to emulate and solve impurity problems, which promotes the utilization of  entanglement-based observables.

Perhaps the most ubiquitous impurity model is that of the Kondo effect, where a magnetic impurity is screened by an electron cloud and forms a so-called many-body Kondo singlet~\cite{kondo_resistance_1964}. Commonly, the effect arises in both strongly correlated materials and mesoscopic quantum devices and is identified via stationary transport signatures~\cite{abrikosov_magnetic_1969,hewson_kondo_1993,jones_low-temperature_1988,glazman_resonant_1988,ng_-site_1988,kawabata_electron_1991}. 
Within NISQ, the emulation of the Kondo effect has been proposed, for example with superconducting qubits circuits~\cite{garcia-ripoll_quantum_2008}. However, such a realization requires new detection schemes, as stationary transport is not applicable within the finite size of the quantum simulators~\cite{dias_da_silva_transport_2008}. Current efforts in this direction focus on the direct characterisation of the amount of entanglement in Kondo singlets
in quantum dots by measuring the dot's entropy~\cite{kleeorin_how_2019,child_entropy_2021} or by quantifying the entanglement in Kondo clouds~\cite{bayat_negativity_2010,bayat_entanglement_2012, lee_macroscopic_2015,eriksson_impurity_2011, kim_universal_2021}.

The more complex the internal structure of a quantum impurity is, the broader the variety of many-body phenomena that can arise. An example of such a structured quantum impurity problem involves an electronic dot-cavity system~\cite{rossler_transport_2015,ferguson_long-range_2017,nicoli_cavity-mediated_2018}. The model's impurity consists of a quantum dot that is coupled to an electronic cavity, i.e., to a discrete set of electronic noninteracting levels, as well as to three electron leads~\cite{ferguson_long-range_2017}. The system resembles a double dot system~\cite{dias_da_silva_zero-field_2006,dias_da_silva_spin-polarized_2013}, where one of the two dots is large and noninteracting, i.e., it is equivalent to a so-called Kondo box~\cite{thimm_kondo_1999} in the large level-spacing limit. The open dot-cavity model was used to describe the transport signatures of a mesoscopic device, that showcased a metal-to-insulator phase transition. The transition was postulated to arise when a dot-cavity
molecular singlet forms and quenches the conducting Kondo effect of the dot~\cite{rossler_transport_2015}. Thus far, only approximate methods were used to describe the dot-cavity phase transition~\cite{ferguson_long-range_2017,dias_da_silva_orbital_2004}.

In this work, we devise measurable entanglement witnesses for the observation of strong correlations in quantum impurities. Specifically, we harness the purity of subparts of the structured impurity as the entanglement witness. As an example for our scheme, we report on an exact study of the dot-cavity system~\cite{rossler_transport_2015,ferguson_long-range_2017,nicoli_cavity-mediated_2018} and its nonlocal entanglement characteristics. We use the purity of the dot and that of the dot-cavity to resolve whether the dot forms a many-body Kondo singlet with its leads or is isolated from the environment by forming a singlet with the cavity. Our analysis relies on a numerical non-perturbative method for the study of quantum impurity systems that combines the strengths of both Numerical Renormalisation
Group (NRG) and Matrix Product States (MPS) formalism~\cite{saberi_matrix-product-state_2008,weichselbaum_variational_2009}. Last but not least, we propose a measurement protocol to detect the purity of the selected subsystem and motivate alternative experimental observables for detecting strong correlations in quantum impurities.

\section{Purity}
\begin{figure}
	\centering\includegraphics[width=8.6cm]{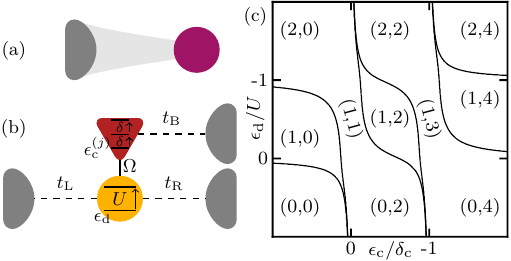}
	\caption{(a) A quantum impurity (purple dot) coupled to a noninteracting environment (grey shape) is a paradigmatic setting for studying the build-up of many-body strong correlations between the impurity and its environment~\cite{ashida_quantum_2020}. (b) Sketch of the dot-cavity (open Kondo box) model~\cite{ferguson_long-range_2017}, cf.~Eq.~\eqref{eq: kondo box model}. A spinful single-level quantum dot (yellow) with energy $\epsilon_\text{d}$ and  on-site interaction $U$ is tunnel-coupled to two leads $\text{L,R}$ with tunneling amplitudes $t_\text{L}, t_\text{R}$, respectively, as well as to an electronic cavity (red) with tunnel-coupling $\Omega$. The cavity is composed of a set of spinful noninteracting levels at energies $\epsilon_\text{c}^{(j)}$ and level spacing $\delta$. It is tunnel-coupled to its own lead with rate $t_\text{B}$. (c) Charge stability diagram of the closed dot-cavity system ($t_i{=}0$, $i{\in}\left\{\text{L},\text{R},\text{B}\right\}$) as a function of the dot level $\epsilon_\text{d}$ and cavity energy offset $\epsilon_\text{c}$. Regions with a different total number of electrons $n_{\text{tot}}{=}\langle n_\text{d}{+}n_\text{c}\rangle$ on the dot-cavity molecule are separated by lines, whereas population within the molecule are marked by $(n_\text{d},n_\text{c})$. The diagram is obtained using exact diagonalisation with Fermi energy $\epsilon_\text{F}{=}0$, cavity level spacing $\delta{=}U$, tunnel-coupling $\Omega{=}0.2U$, and the cavity Hilbert space is truncated to include two spin-degenerate levels. }\label{fig:model}
\end{figure}

We are interested in the entanglement of a quantum impurity with its environment, see Fig.~\ref{fig:model}(a). As a witness for the entanglement, we will use the purity of the impurity together with the purity of its subparts. In a bipartite closed system $A\otimes B$, the purity $\mathcal{P}_A$ of a subsystem $A$ (e.g., of the impurity) reads
\begin{equation}\label{eq: purity bipartite system text}
	\mathcal{P}_A = \Tr\left[\rho_A^2\right] \ ,
\end{equation}
where  $\rho_A^{\phantom 2}{=}\Tr_B (\rho)$ is the reduced density matrix of subsystem $A$ obtained by tracing out subsystem $B$ (the environment). The purity is a witness for entanglement: when $A$ is decoupled from its environment, then $\mathcal{P}_A{=}1$; if $A$ is fully mixed with the environment, then $\mathcal{P}_A{=}1/n$, where $n$ is the size of the Hilbert space of $A$, e.g., $1/4$ for a single spinful electron site, and $1/16$ for a system with two spinful electron sites. Crucially, when a singlet forms between $A$ and $B$ the purity is $\mathcal{P}_A{=}1/2$. Note that other entanglement measurements can be equivalently evaluated,
and would result in the same physical picture of the model~\cite{amico_entanglement_2008}. Yet, because the purity is calculated based on the density operator of the subsystem only, see Appendix~\ref{appendix: purity in MPS}, we show here that it engenders significantly-low computational costs and becomes amenable to experimental observation in small impurity systems.  

\section{Model}\label{sec: model}
The effective Hamiltonian of the dot-cavity system reads~\cite{rossler_transport_2015,ferguson_long-range_2017} [cf.~Fig.~\ref{fig:model}(b)]
\begin{align}
	H & =   H_{\text{dot}} + H_{\text{cav}} + H_{\text{coupl}} + H_{\text{leads}} + H_{\text{tun}}\ .
	\label{eq: kondo box model}
\end{align}
The dot Hamiltonian
\begin{equation}\label{eq: dot hamiltonial}
\Hdot=\sum_\sigma^{\phantom \dag}\epsilon_\text{d}^{\phantom\dag}n_{\text{d}\sigma}^{\phantom\dag}+U n_{\text{d}\uparrow}^{\phantom\dag}n_{\text{d}\downarrow}^{\phantom\dag}
\end{equation}
describes an Anderson impurity~\cite{anderson_localized_1961} with a spin-degenerate electron level at energy $\epsilon_{\text{d}}$ and electron-electron charging energy $U$. Here, $n_{\text{d}\sigma}^{\phantom\dag}$ denotes the dot level's occupation number with spin $\sigma{\in}\{\uparrow,\downarrow\}$.  The cavity Hamiltonian is 
\begin{equation}\label{eq: cav hamiltonian}
H_{\text{cav}}=\sum_{j\sigma}^{\phantom\dag}\epscj\fdag_{j\sigma} \fpdag_{j\sigma} \ ,
\end{equation}
where the index $j$ labels spin-degenerate electron levels with equally-spaced energies $\epscj{=}\epsilon_{\text{cav}}^{\phantom\dag}{+}j\delta$ with spacing $\delta$, and  $\fdag_{j\sigma}$ $(\fpdag_{j\sigma})$ is the fermionic cavity creation (annihilation) operator of the $j$th level. Note that due to the screening within the large spatial extent of mesoscopic electronic cavities~\cite{rossler_transport_2015,ferguson_long-range_2017}, no electron-electron repulsion term is introduced in $H_{\text{cav}}$. 
The dot and cavity are tunnel-coupled 
\begin{equation}
H_{\text{coupl}}^{\phantom\dag}=\sum_{j\sigma}^{\phantom\dag}\Omega_j^{\phantom\dag} \ddag_{\sigma}\fpdag_{j\sigma}+ \text{H.c.} \ ,
\end{equation}
with energy-dependent coupling amplitudes $\Omega_j$.
The environment is composed of three leads
\begin{equation}\label{eq: leads}
H_{\text{leads}}^{\phantom\dag}=\sum_{k\ell\sigma}^{\phantom\dag}\epsilon_{k\ell}^{\phantom\dag}\cdag_{k\ell\sigma}\cpdag_{k\ell\sigma}+ \sum_{k\sigma}^{\phantom\dag}\epsilon_{k\text{B}}^{\phantom\dag}\cdag_{k\text{B}\sigma}\cpdag_{k\text{B}\sigma}\ ,
\end{equation}
where we denote $c_{k\ell\sigma}^{\phantom\dag}$ ($\cdag_{k\ell\sigma})$ the fermionic annihilation (creation) of an electron with momentum $k$ and spin $\sigma$ in the leads left and right to the dot;  $\ell{\in}\{\text{L},\text{R}\}$. The $c_{k\text{B}\sigma}^{\phantom\dag} \ (\cdag_{k\text{B}\sigma})$ operator acts on the cavity lead and is defined analogously. The tunneling Hamiltonian term $H_{\text{tun}}^{\phantom d}{=}H^{\text{dot}}_{\text{tun}}{+} H^{\text{cav}}_{\text{tun}}$ 
couples the closed dot-cavity system to the leads. We assume that the dot is coupled to the left and right leads, with energy-independent tunneling amplitudes $\tdl$; 
\begin{equation}\label{eq: dot-leads tunneling}
H_{\text{tun}}^{\text{dot}}=\sum_{\ell  k\sigma}\tdl\ddag_\sigma \cpdag_{k\ell\sigma}+\text{H.c.} \ .
\end{equation}
Similarly, the cavity is tunnel-coupled to its own lead
\begin{equation}\label{eq: cav-lead tunneling}
H_{\text{tun}}^{\text{cav}}=\sum_{j k\sigma}^{\phantom\dag}\tcB\fdag_{j\sigma} \cpdag_{k\text{B}\sigma}+\text{H.c.} \ ,
\end{equation}
with an energy-independent tunneling amplitude $t_{\text{B}}^{\phantom\dag}$~\footnote{Due to the fact that no Fano signatures are observed in the experiment~\cite{rossler_transport_2015,ferguson_long-range_2017}, and unlike the case studied in Ref.~\cite{dias_da_silva_conductance_2017}.}. 

As discussed in Ref.~\cite{ferguson_long-range_2017}, the closed dot-cavity system, i.e., Eq.~\eqref{eq: kondo box model}  with $t_{\text{L}}^{\phantom\dag}, t_{\text{R}}^{\phantom\dag}, t_\text{B}^{\phantom\dag}{=}0$, forms an ``artificial molecule'' impurity. Specifically, as a function of dot and cavity energies (set experimentally by tuning voltage gates), it exhibits a charge stability diagram with regions dominated by Coulomb or exchange blockade, separated by resonance lines where the particle number on the molecule is not conserved, see Fig.~\ref{fig:model}(c). The exchange blockade region involves the formation of a dot-cavity singlet for sufficiently-large level spacing in the cavity $\delta {\gg}12 \Omega_j^2 /U$.

\section{Methods}
Coupling the dot-cavity molecule to its leads can change the occupation of the molecule and transport through it. In experiments~\cite{rossler_transport_2015}, Kondo transport is observed, separated by exchange-blockade regions whenever the cavity couples strongly to the dot's electron spin. In Ref.~\cite{ferguson_long-range_2017}, a strong hybridisation between dot and cavity was assumed alongside a perturbative treatment with respect to the rest of the electronic environment. This observation led to the conjecture that the blockade regions form due to the formation of a spin-singlet between dot and cavity. Yet, it was not known whether this singlet is decoupled or influenced by the hybridisation with the environment. Here, instead, we treat the system-environment couplings exactly with the NRG-MPS method for quantum impurity systems~\cite{saberi_matrix-product-state_2008,weichselbaum_variational_2009}.

According to the NRG-MPS method for quantum impurity systems~\cite{saberi_matrix-product-state_2008,weichselbaum_variational_2009}, we transform the infinite environment of the dot-cavity system~[Eqs.~\eqref{eq: leads}-\eqref{eq: cav-lead tunneling}] in 1D Wilson chains. In the equilibrium configuration (zero bias voltage $\mu_\text{L}{=}\mu_\text{R}{=}\mu_\text{B}$ across the device), we can combine the identical left and right leads using the transformation  $\cpdag_{k\text{P}\sigma}{=}\alpha\left(\tdL\cpdag_{k\text{L}\sigma}{+}\tdR\cpdag_{k\text{R}\sigma} \right)$,
whose orthogonal complementary combination  $\cpdag_{k\text{M}\sigma}{=}\alpha\left(\tdR\cpdag_{k\text{L}\sigma}{-}\tdL\cpdag_{k\text{R}\sigma} \right)$
fully decouples from $H_{\text{dot}}$ and the rest of the system~\cite{glazman_resonant_1988}. The prefactor $\alpha$ is a normalisation constant. The leads' Hamiltonian, Eq.~\eqref{eq: leads}, thus respectively transform into
\begin{equation}
	H_{\text{tun}}^{\text{dot}}\mapsto\sum_{k\sigma}^{\phantom\dag}\td\ddag_\sigma \cpdag_{k\text{P}\sigma}+\text{H.c.}
\end{equation}
and 
\begin{equation}
H_\text{leads}^\text{dot} \mapsto\sum_{ks\sigma}^{\phantom\dag}\epsilon_{k}^{\phantom\dag}c_{ks\sigma}^{\dag}\cpdag_{ks\sigma}+ \text{H.c.} \ ,
\end{equation}
where  $s{\in}\{\text{P},\text{M}\}$. As the transformed M lead is decoupled from the system, we can henceforth neglect it.

\begin{figure}
	\centering\includegraphics[width=8.6cm]{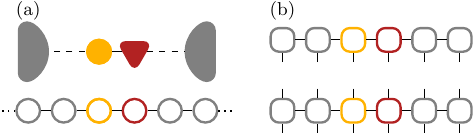}
	\caption{NRG-MPS reshaping of the dot-cavity Hamiltonian. (a) The reshaped dot-cavity model [cf.~Eq.~\ref{eq: kondo box model} and Fig.~\ref{fig:model}(b)] is transformed into a 1D chain using Wilson's transformation~\cite{wilson_renormalization_1975, bulla_numerical_2008}. Open circles denote sites along the resulting chain with colors grey, yellow and red matching the leads, dot and 
		cavity, respectively. (b) For numerical calculations, we characterise states and operators along the chain using the MPS formalism~\cite{schollwock_density-matrix_2011}, denoted by open rounded squares.}\label{fig:nrg-mps technique}
\end{figure}

As a next step, we map the remaining $\text{P}$ lead and its tunnel-coupling to a 1D semi-infinite Wilson chain~\cite{wilson_renormalization_1975}, see Fig.~\ref{fig:nrg-mps technique}(a),
\begin{equation}\hat{H}^{\text{dot}}_{\text{chain}} =\sum_{\sigma}^{\phantom\dag}t_\text{d}^{\phantom\dag}d_\sigma^\dag c_{0\sigma}^{\phantom\dag}+\sum_{n\sigma}^\infty \tdn\cdag_{n\sigma}c_{n+1\sigma}+\text{H.c.} \ , 
\end{equation}
where $c_{n\sigma}$ is the annihilation operator of a fermion with spin $\sigma$ at site $n$ of the P-chain. Similarly, we transform the terms including the cavity lead into a Wilson chain 
\begin{equation}
\hat{H}^{\text{cav}}_\text{tun}=\sum_{\sigma}^{\phantom\dag} t_{\text{c}}^{\phantom\dag}f_\sigma^\dag b_{0\sigma}^{\phantom\dag}+\sum_{n\sigma}^\infty \tcn b^\dag_{n\sigma}b_{n+1\sigma}^{\phantom\dag}+\text{H.c.} \ ,
\end{equation}
where $b_{n\sigma}$ is the annihilation operator in the cavity lead defined analogously to $c_{n\sigma}$. From the Wilson transformation~\cite{wilson_renormalization_1975}, the nearest-neighbour hopping coefficients along the chain are exponentially decaying in the chain site number $t_{\text{d}n}{=}t_{\text{c}n}{\propto}\Lambda^{-n/2}$ for $n{\gg}1$.

An impurity problem with environments that are transformed into Wilson chains is commonly studied with the NRG technique~\cite{wilson_renormalization_1975,krishna-murthy_renormalization-group_1980,krishna-murthy_renormalization-group_1980-1}. In our work, we instead calculate equilibrium
properties of a finite chain with a variational algorithm for MPS
\cite{saberi_matrix-product-state_2008,weichselbaum_variational_2009}. As sites beyond some point in the Wilson chain become marginal~\cite{bulla_numerical_2008}, we truncate both the dot and cavity leads at a finite length $N$ (here $N{=}40$), obtaining a 1D finite chain that approximates the open dot-cavity system. Nevertheless, recall that each chain
site has a $d{=}4$ dimensional Hilbert space with empty $\ket{0}$,
full $\ket{\uparrow\downarrow}$, or singly occupied $\ket{\uparrow}$, $\ket{\downarrow}$ basis states. Therefore, the number of coefficients required to characterise
a pure state of the chain grows as $d^N$, hitting the current computational limits at relatively small $N$. On this
account, we work with MPS to characterise states and
operators [Fig.~\ref{fig:nrg-mps technique}(b)], which allows for a controlled truncation of the number of required coefficients based on the
entanglement information between different chain sites.
This truncation harnesses a Singular Value Decomposition (SVD) compression that caps the maximal number of Schmidt values describing each link in the chain,
dubbed bond dimension D. Only thus, are we able to
reach chains of length $N_{\text{tot}}{>}80$ sites~\footnote{in the study of a single-channel Anderson impurity, so-called ``unfolding'' of the electron chain resulted in a considerable performance increase~\cite{saberi_matrix-product-state_2008}. For the multi-channel structured impurity studied here, we obtain better convergence with ``folded leads''.}.

We analyse the full system at equilibrium, i.e., we set a zero bias voltage $\mu_\text{L}{=}\mu_\text{R}{=}\mu_\text{B}$ across the device, where $\mu_i$ denotes the chemical potential of the $i$th lead. Even though the NRG-MPS method does not treat the impurity-environment coupling perturbatively, its performance depends on the system parameters: as the impurity-environment coupling becomes smaller, the required resolution to find the Kondo ground state increases. In the following, we establish that the purity is a good witness for observing the formation of the Kondo singlet. Hence, per system parametrisation, we systematically explore convergence of our variational ground-state search by following the dot's purity $\mathcal{P}_{\text{d}}$ as a function of the NRG-MPS numerical truncation knobs, see Appendix~\ref{appendix: convergence NRG-MPS}.
\begin{figure}
	\centering \includegraphics[width=8.6cm]{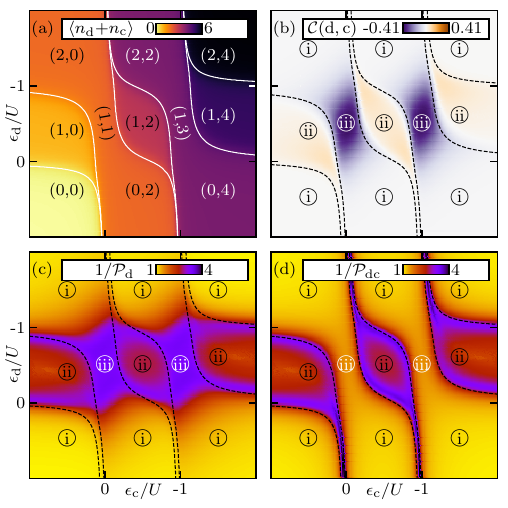}
	\caption{Dot-cavity observables [cf.~Eq.~\eqref{Eq: correlation function}] and purity [cf.~Eq.~\eqref{eq: purity bipartite system text}] calculated with the NRG-MPS approach as a function of the dot energy $\epsilon_\text{d}$ and the cavity energy $\epsilon_\text{c}$. (a) Occupation of the open dot-cavity artificial molecule. We denote $(\langle n_{\text{d}}\rangle,\langle n_\text{c}\rangle)$ the dot and cavity occupation. Solid lines are superimposed from the closed system calculation [cf.~Fig.~\ref{fig:model}(c)]. (b) Dot-cavity spin-spin correlation, cf. Eq.~\eqref{Eq: correlation function}. (c) Inverse of the purity of the dot $1/\mathcal{P}_\text{d}$ and (d) of the dot-cavity $1/\mathcal{P}_\text{dc}$ molecule. The dashed lines in (b)--(d) indicate the contour of the numerical results of (a). The coupling parameters of the system are as in Fig.~\ref{fig:model}(c), with $t_{\text{d}}{=}0.25U$, $t_{\text{c}}{=}0.18U$, $\rho{=}U$, and constant leads' density of states $\mathcal{D}_0{=}1/4U$. We set MPS bond dimension $D{=}100$ and a Wilson chain length of $N{=}40$ for both the dot and cavity lead. Regions (i)-(iii) mark distinct restructuring in the many-body states, cf. Section~\ref{sec: results}.}
	\label{fig:nrg-mps results}
\end{figure} 
Furthermore, we observe that the physics of the system only depends on the presence of an unpaired cavity electron and is fully equivalent between the different cavity levels $\epsilon_\text{c}{=}\epsilon_\text{cav}{+}j\delta$ in the experimentally-relevant regime where $\delta{\gg} 12\Omega^2/U$, see Appendix~\ref{appendix: truncatio cavity level}

\section{Results}\label{sec: results}
Having obtained the converged ground state of the whole system (impurity plus environment) per parametrisation, we start by revisiting the charge stability diagram, see Fig.~\ref{fig:nrg-mps results}(a). The occupation $n_\text{tot}{=}\langle n_\text{d}{+}n_\text{c}\rangle$ of the dot-cavity subsystem (evaluated on the obtained ground state) qualitatively matches that obtained by the exact diagonalisation treatment of the closed dot-cavity system. Crucially, we observe a region competing with the dot's Coulomb blockade regimes, where due to the dot-cavity tunnel-coupling, the cavity has an odd occupation. This hints towards the fact that in this region a dot-cavity spin-singlet forms (exchange-blockade) also in the open system. Quantitatively, we observe that the coupling to the environment shifts the charge instability lines to slightly higher energies, as expected from self-energy renormalisation of the closed systems' levels by high-order coupling to the environment~\cite{ferguson_long-range_2017}. 

To verify the formation of a dot-cavity spin-singlet and the existence of an exchange blockade region, we analyse the spin configuration of the dot-cavity molecule. To this end, we define and calculate their spin-spin correlation
\begin{align}
	\mathcal{C}\left(\text{d},\text{c}\right) = \frac{1}{2}\sum_{\sigma}^{\phantom\dag}\langle n_{\text{d}\sigma}\rangle\langle n_{\text{c}\bar{\sigma}}\rangle - \langle n_{\text{d}\sigma}n_{\text{c}\bar{\sigma}}\rangle \ ,
	\label{Eq: correlation function}
\end{align}
which returns 1 (-1) when the spins on the dot and cavity are aligned (anti-aligned). We plot the resulting $\mathcal{C}\left(\text{d},\text{c}\right)$ in Fig.~\ref{fig:nrg-mps results}(b) and observe three main phases (i)-(iii). (i) When the dot has an even occupation, no dot-cavity spin-spin correlations appear. (ii) When the dot has an odd occupation and the cavity has an even occupation, we observe weak spin-spin alignment. In this region, we expect that the dot anti-aligns with the lead electrons (Kondo singlet). The ensuing strong dot-lead correlation leaves space for the cavity electrons (holes) to partially redistribute into the leads in the full (empty) cavity level configuration. The spin of these residual particles tends to align with the spin of the electron occupying the dot, see Appendix~\ref{appendix: dot-cavity alignment}. (iii) In the (1,1) occupation region, we observe strong spin anti-alignment behaviour, in agreement with the closed system exact diagonalisation results. The latter clearly confirms the hypothesised formation of a molecular dot-cavity singlet~\cite{rossler_transport_2015,ferguson_long-range_2017}. 

The results presented so far do not identify signatures of strong coupling to the environment. To observe these strong correlations, we propose to use quantum information inspired observables, namely the dot and dot-cavity purities $\mathcal{P}_\text{d},\ \mathcal{P}_{\text{dc}}$, see Figs.~\ref{fig:nrg-mps results}(c)~and~(d), and cf.~Eq.~\ref{eq: purity bipartite system text}. Combining the information from both purities, we can deduce much about the correlation in the full system. For example, in region (i) both $\mathcal{P}_{\text{d}}{\to}1$ and $\mathcal{P}_{\text{dc}}{\to}1$, i.e., the electron on the dot is decoupled from all lead and cavity electrons, and the combined dot-cavity electrons are also decoupled from the lead electrons. This decoupling highlights that in region (i) the many-body ground state does not show strong correlations between the impurity and the environment. Similarly, in 
region (iii), the dot-cavity ``molecule'' decouples from the leads $(\mathcal{P}_{\text{dc}}{\to}1)$. Yet a $\mathcal{P}_{\text{d}}{\approx}1/2$ attests that the dot electron is in a spin-singlet configuration with its cavity counterpart. Crucially, in region (ii), $\mathcal{P}_{\text{d}}{\approx}\mathcal{P}_{\text{dc}}{\approx}1/2$ marks that the dot forms a singlet, but that this singlet is not affected by the inclusion of the cavity's Hilbert space. Here, we have a clear evidence for the formation of a Kondo singlet between the dot and lead electrons. Note that our environment model includes 2D fermionic baths via Wilson chains and does not consists of tailored 1D environments as considered in previous works~\cite{bayat_negativity_2010,lee_macroscopic_2015}. Additionally, we model the quantum dot with an Anderson Impurity Hamiltonian instead of the Kondo Hamiltonian considered in Ref.~\cite{weichselbaum_variational_2009}, i.e., we did not a priori assume the formation of Kondo effects. Thereby, our method does not merely verify the Kondo singlet formation for the Anderson model; it additionally fully characterises the phase transitions to other many-body entanglement regimes in the structured impurity of the open Kondo box. 

\section{Measurement protocol} 
\begin{figure}[h]
	\centering\includegraphics[width=8.6cm]{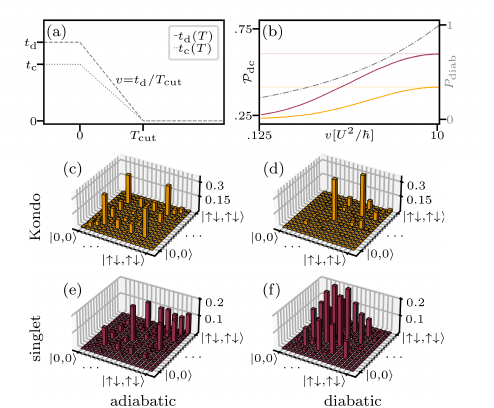}
	\caption{Procedure and results of the measurement protocol. (a) The dot and cavity leads are linearly detached within a time interval $T_\text{cut}$ with velocity $v{=}t_{\text{d}}/T_\text{cut}$. (b) Purity of the dot-cavity impurity measured after a detachment with velocity $v$ for a singly occupied dot $\epsilon_\text{d}{=}{-}U/2$ in the Kondo ($\epsilon_\text{c}{=}{-}U/2$, yellow line) and dot-cavity singlet ($\epsilon_\text{c}{=}0$, red line) regimes. Horizontal transparent lines mark the numerical value obtained by the ideal trace operation, cf.~\ref{fig:nrg-mps results}(d). The dashed (gray) line marks the estimated diabatic transition probability, cf.~Eq.~\eqref{eq: probability diabatic}. (c)-(f) Full state tomography of the dot-cavity impurity in the Kondo regime [(c) adiabatic with $v{=}0.125U^2/\hbar$ and (d) diabatic with $v{=}10U^2/\hbar$]. (e) and (f) show the same comparison for the dot-cavity singlet case. The other coupling parameters of the system are as in Fig.~\ref{fig:nrg-mps results}(c).}
	\label{fig: tomography results}
\end{figure} 
To emulate the action of the partial trace [cf.~Eq.~\eqref{eq: purity bipartite system text}], our entanglement witness can be experimentally measured using a fast (diabatic) gate pulse that detaches the impurity from its environment. After this pulse, the purity of the subsystem can be read out using tomography techniques, that are commonly harnessed in  superconducting and silicon spin qubit devices~\cite{cogan_complete_2020,yang_operation_2020,leon_bell-state_2021,gachter_single-shot_2022}. Hence, in the following, we focus on providing an estimation for the gate pulse frequency necessary for a diabatic lead detachment. We detach the leads within the time interval $T_{\text{cut}}$ by linearly decreasing the dot-lead and cavity-lead coupling coefficients $t_\text{d}, t_\text{c}$, see Fig.~\ref{fig: tomography results}(a). We simulate this protocol using Time Evolving Block Decimation~\cite{dias_da_silva_transport_2008,paeckel_time-evolution_2019} on the dot-cavity MPS state for both the Kondo and singlet cases. Following the detachment, tomography on the dot-cavity subsystem is performed to evaluate its purity, see Fig.~\ref{fig: tomography results}(b)-(f). The resulting purity deviates from the ideal case when  $v{=}t_\text{d}/T_\text{cut}$ is insufficiently large. However, with increasing velocity we obtain a very good agreement. 

Using the Landau-Zener formula~\cite{landau1932theorie,zener_non-adiabatic_1932}, 
\begin{equation}
	P_\text{diab} = \exp\left[-2\pi J^2\left(\hbar\left|\frac{dq}{dt}\frac{\partial}{\partial q} \Delta E\right|\right)^{-1}\right] \ ,
	\label{eq: probability diabatic}
\end{equation}
the performance of our protocol improves when the probability for a diabatic transition to occur $P_{\text{diab}}{\to}1$, see Fig.~\ref{fig: tomography results}(b). Here, 
\begin{equation}
\Delta E\approx \sqrt{\pi t_{\text{d}}^2\mathcal{D}_0U/2}\exp\left[-U/\left(8t_{\text{d}}^2\mathcal{D}_0\right)\right]
\end{equation}
is the energy gap between the many-body ground and first excited state of the system~\cite{schrieffer_relation_1966}, $J{\sim}\Delta E$ is the coupling amplitude between these states, and $q{=}t_\text{d}(T)$ is the detachment perturbation parameter. 

Using Eq.~\eqref{eq: probability diabatic}, we estimate the minimal gate pulse frequency for a sufficiently diabatic lead detachment. In the experiment, $U{\approx}700\si{\micro eV}$~\cite{rossler_transport_2015}. Assuming a typical semiconductor device bandwidth of $W{=}0.1\si{eV}$ and lead tunnel coupling  
\begin{equation}
    \Gamma_\text{S}+\Gamma_\text{D}\equiv 2\pi\left|t_\text{d}\right|^2\mathcal{D}_0=80\si{\micro eV} \ ,
\end{equation} 
the diabatic probability is $P_\text{diab}{>}0.92$ for $T_\text{cut}{=}1\si{ns}$. 
Thus, with a pulse in the $\si{GHz}$ regime, a diabatic detachment of the leads from the dot-cavity impurity is possible. 
\section{Conclusion}
Our work demonstrates that the purity of different subparts can be harnessed to study complex quantum impurity systems and highlights the applicability of quantum information based observables for studying strongly correlated systems. Employing minimal and sufficient entanglement measure (such as the purities proposed here) in experiments is an important next step for the quantum simulation field, where tomography on small subsystems is readily accessible. Additionally, we showcase three important developments: (a) an experimental measurement protocol of the purity, (b) the power of NRG-MPS approaches for solving multi-channels quantum impurity problems as insinuated in Refs.~\cite{saberi_matrix-product-state_2008,weichselbaum_variational_2009}, and (c) the conclusive demonstration of the physics behind the dot-cavity metal-to-insulator transition. Future work will explore the generalisation of entanglement quantification~\cite{ma_symmetric_2022, carisch_mixed_2022} in
out-of-equilibrium and mixed state quantum impurity systems~\cite{lotem_renormalized_2020}.\\

\begin{acknowledgements} We thank G. Blatter, T. Ihn, K. Ensslin, M. Goldstein, C. Carisch, and J. del Pino for illuminating discussions and acknowledge financial support from the Swiss National Science Foundation (SNSF) through project 190078, and from the Deutsche Forschungsgemeinschaft (DFG) - project number 449653034. Our numerical implementations are based on the ITensors \textsc{Julia} library~\cite{fishman_itensor_2020}.\\
\end{acknowledgements} 

\appendix
\section{Convergence behaviour of the NRG--MPS technique}\label{appendix: convergence NRG-MPS}
\begin{figure}
	\centering\includegraphics[width=8.6cm]{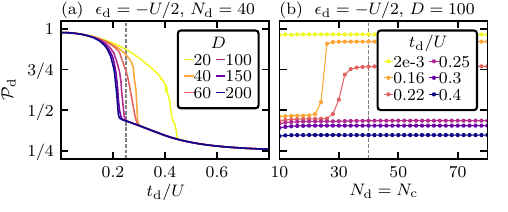}
	\caption{Convergence analysis for $\mathcal{P}_\text{d}$ for a singly occupied dot, and a cavity level located far away from resonance $\epsilon_\text{c}{=}U$ (a) as a function of dot-lead tunnel coupling $t_\text{d}$ and MPS bond dimension $D$ for a fixed Wilson chain length $N{=}40$, and (b) as a function of the Wilson $N$ and tunnel-coupling $t_\text{d}$ for a fixed bond dimension $D{=}100$.  
	}\label{fig: MPS convergence analysis}
\end{figure}

The quality of a Matrix Product State (MPS) or operator (MPO) representation is affected by the extent of the singular value decomposition (SVD) truncation, namely by choosing the bond dimension $D$. There is no general rule for determining the minimal $D$ that ensures a converged physical result. Hence, trials with different bond dimensions are performed. Convergence properties of physical observables might indicate a minimal bond dimension that is still able to capture the physics of the system. Similarly to the MPS bond dimension, also the length of the Wilson chain $N$ must be sufficiently large in order to observe the formation of a Kondo singlet. We checked the results of the dot-cavity model for convergence. For example, in Fig.~\ref{fig: MPS convergence analysis}(a), we observe a strong dependence of the dot purity $\mathcal{P}_\text{d}$ on the bond dimension. At high dot-lead coupling $t_\text{d}/U{\gg}0$, the dot fully mixes with the environment, and $\mathcal{P}_\text{d}{\approx} 1/4$ is minimal for any bond dimension. For lower coupling strengths, the purity gradually increases to approximately $1/2$, marking the formation of a Kondo singlet. The jumps in $\mathcal{P}_\text{d}$ observed in the $0.2U{<} t_\text{d}{<}0.3U$ range showcase the aforementioned limitation of the MPS method: a finite MPS bond dimension is not sufficient to capture the physics of the system. Indeed, a purity value indicating the presence of a dot-lead Kondo singlet is expected for any $t_\text{d}$ in the low coupling regime.

In Fig.~\ref{fig: MPS convergence analysis}(b), we repeat the convergence analysis of the $\mathcal{P}_\text{d}$ for varying Wilson chain length $N$ for both the dot and cavity leads. We see that the weaker the dot-lead tunnel-coupling is, the longer $N$ has to be to capture the Kondo-associated entanglement features established in the system. This is a known requirement in NRG since the Wilson chain relates to the energy resolution of the lead. A finite chain truncates the information corresponding to coupling with lead electrons near the Fermi level. These low-energy electrons are the ones dominating the hybridisation with the dot in the low tunnel-coupling regime.

Following our calibration, in the results presented in the main text, we set the dot-lead tunneling amplitude at $t_\text{d}{=}0.25U$, as well as use MPS bond dimension $D{=}100$ and an NRG chain length $N{=}40$ for both the dot and cavity leads, i.e., where the purity value is close to 1/2, and the expected dot-lead Kondo singlet is detected.
\section{Truncation to a single cavity level}\label{appendix: truncatio cavity level}
\begin{figure}
	\centering\includegraphics[width=8.6cm]{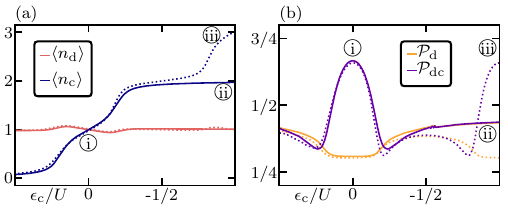}
	\caption{Comparison between single-level (solid line) and two-level (dotted line) cavities at $\epsilon_{\text{d}}{=}{-}U/2$. (a) The dot and cavity occupation $\langle n_{\text{d}}\rangle$ and $\langle n_{\text{c}}\rangle$, respectively. (b) The purity of the dot and dot-cavity subsystems $\mathcal{P}_{\text{d}},\mathcal{P}_{\text{dc}}$, respectively. We use here the parameters as in Fig.~\ref{fig:nrg-mps results}.}  \label{fig: single cavity level}
\end{figure}

In Sec.~\ref{sec: model} of the main text, we define and discuss the cavity Hamiltonian $
H_{\text{cav}}$, whose spin-degenerate electron levels $\epscj{=}\epsilon_{\text{cav}}^{\phantom\dag}{+}j\delta$ have constant spacing $\delta$, cf.~Eq.~\eqref{eq: cav hamiltonian}. For the presented results, however, we suffice to include only a two-level cavity. In this Section, we check the validity of this approximation. In Fig.~\ref{fig: single cavity level}, we compare the outcome of our VMPS analysis between a dot-cavity system with a single-level cavity and a two-level one. We perform the comparison for a dot located at $\epsilon_\text{d}{=}{-}U/2$, where the competition between Kondo and dot-cavity singlet formation occurs.

In Figs.~\ref{fig: single cavity level}(a)~and~(b), we show that for $\epsilon_{\text{c}}{>}{-}U/2$ the single-level cavity exhibits the same physics as the two-level cavity. As discussed in Sec.~\ref{sec: results} of the main text, at $\epsilon_{\text{c}}{\approx}0$ [cf. region (i)], the cavity is approximately occupied by a single electron and the dot-cavity molecule decouples from the environment [$\mathcal{P}_{\text{dc}}{\to}1$]. For a cavity approximately empty or full, the purities are $\mathcal{P}_{\text{d}}{\approx}\mathcal{P}_{\text{dc}}{\approx}1/2$. Hence, we conclude that the cavity does not impact the many-body strong hybridisation between the dot and its environment, namely the dot-lead Kondo singlet.

A different picture emerges for $\epsilon_{\text{c}}{<}{-}U/2$. In region (ii), for the single-level cavity case, the occupation $\langle n_{\text{c}}\rangle{=}2$ saturates and the system remains in the dot-lead singlet configuration. For the two-level case, a third electron occupies the cavity $\langle n_{\text{c}}\rangle{>}2$. At $\epsilon_{\text{c}}{\approx}{-}U$, $\mathcal{P}_{\text{d}}$ and $\mathcal{P}_{\text{dc}}$ of the two-level cavity [cf. region (iii)] are equal to the $\mathcal{P}_{\text{d}}$ and $\mathcal{P}_{\text{dc}}$ at $\epsilon_{\text{c}}{\approx}0$ [cf. region (i)]. Thus, in the $\langle n_{\text{c}}\rangle{\approx}3$ configuration the unpaired cavity electron couples to the dot, exactly as in the $\langle n_{\text{c}}\rangle{\approx}1$ configuration, and establishes a molecular dot-cavity singlet quenching the Kondo effect. 

With these results, we conclude that the physics in the multi-level cavity case only depends on the presence of an unpaired electron in the cavity, regardless of the number of occupied cavity levels.  Thus, it is sufficient to consider a two-level cavity for characterising the many-body physics of the electronic dot-cavity impurity.
\section{Dot-cavity alignment in the Kondo regime}\label{appendix: dot-cavity alignment}
\begin{figure}
	\centering\includegraphics[width=8.6cm]{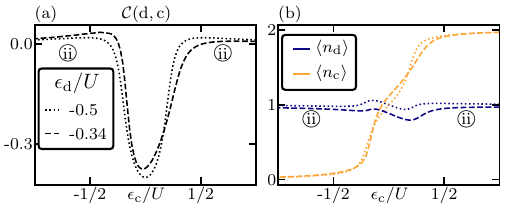}
	\caption{(a) Dot-cavity spin correlation $\mathcal{C}(d,c)$ and (b) dot and cavity occupation [$\langle n_{\text{d}}\rangle$ (blue lines) and $\langle n_{\text{c}}\rangle$ (orange lines), respectively] as a function of the cavity level $\epsilon_\text{c}$. The dot energy level is taken at $\epsilon_{\text{d}}{=}{-}U/2$ (dotted lines) and $\epsilon_{\text{d}}{=}{-}0.34U$ (dashed lines). The rest of the parameters are as in  Fig.~\ref{fig:nrg-mps results}.} \label{fig: cavity leaving outside}
\end{figure}

As discussed in Sec.~\ref{sec: results} of the main text [cf. Fig.~\ref{fig:nrg-mps results}, region (ii)], whenever the dot and lead form a Kondo singlet, we observe a limited dot-cavity spin alignment, see also Fig.~\ref{fig: cavity leaving outside}(a), region (ii). This effect coexists with the strong dot-lead hybridisation and implies that the cavity occupation is not perfectly quantised. A necessary condition for the dot and cavity spin to align is that the cavity is not perfectly full (empty) for positive (negative) $\epsilon_{\text{c}}$. Such behaviour is corroborated by our calculation of $\langle n_{\text{c}}\rangle$, see Fig.~\ref{fig: cavity leaving outside}(b). The cavity electrons (holes) could redistribute into the leads, the dot, or both.

In Fig.~\ref{fig: cavity leaving outside}(b), we observe that the dot occupation is constant in the (ii) region. Thus, in the Kondo dot-lead configuration, the cavity electrons (holes) partially redistribute into the leads in the full (empty) cavity configuration. In the Kondo singlet, a continuous second-order scattering of lead electrons near the Fermi level occurs. Thus, for a dot electron in the $\sigma$ spin configuration, a lead electron in $\bar{\sigma}$ scatters in and out of its energy level~\cite{hewson_kondo_1993}, leaving the latter free for a fraction of time. When the cavity is tuned near the Fermi level, the $\bar{\sigma}$ cavity electron (hole) partially redistributes in the free lead level. The $\bar{\sigma}$ cavity electron (hole) level is therefore partially empty. Hence, dot and cavity showcase a marginal spin-spin alignment behaviour. Our results show that the NRG-MPS method is capable of identifying such higher-order processes.
\section{Purity calculation in MPS formalism}\label{appendix: purity in MPS}
In this Section, we show how to readily calculate the purity of a subsystem using a MPS formalism. The density operator $\rho_A$ of a subsystem $A$ is defined as
\begin{align}
	\rho_A & = \text{Tr}_B(\rho_{AB}) = \displaystyle\sum_{k}\left(I_A\otimes\bra{\sigma_k}_B\right)\rho_{AB}\left(I_A\otimes\ket{\sigma_k}_B\right) \ ,
\end{align}
where $\rho_{AB}$ is the density operator of the full system $A\otimes B$,  $I_A$ is the identity operator on $A$, and $\left\{\ket{\sigma_k}_B\right\}$ is an orthonormal basis of $B$. In our work, we deal with density matrices of a 1D discrete system (chain), whose pure state can be written in Fano form
\begin{equation}
	\label{Eq: pure state}
	\ket{\psi} = \sum_{\vecsig}c_{\sigma_1,\sigma_2\ldots,\sigma_N}\ket{\sigma_1}\otimes\ket{\sigma_2}\otimes\ldots\otimes\ket{\sigma_N} \ ,
\end{equation}
where $\left\{\ket{\sigma_i}\right\}$ is a basis of the Hilbert space of the $i$th chain site, and $c_{\sigma_1,\sigma_2\ldots,\sigma_N}$ are the amplitudes of the specific tensorial basis state.
The reduced density operator of, e.g., the subsystem composed of the first chain reads
\begin{align}
	\label{Eq: App density matrix single dot}
	\rho_{1} & = Tr_{env}(\rho) \\
	& = \sum_{\sigma_2,\ldots,\sigma_N} \left(I_1\otimes\bra{\sigma_2\ldots\sigma_N} \right)\ket{\psi}\bra{\psi}\left(I_1\otimes\ket{\sigma_2\ldots\sigma_N}\right) \ ,
\end{align}
where with $env$ we indicate the environment $\ket{\sigma_2}{\otimes}{\ldots}\ket{\sigma_N}{\equiv}\ket{\sigma_2{\ldots}\sigma_N}$ and $I_1$ the identity operator of the $1^{\rm st}$ site. In MPS formalism, the pure state as in Eq.~\eqref{Eq: pure state} is expressed as
\begin{equation}
	\ket{\psi} = \sum_{\vecsig}M^{\sigma_1}\ldots M^{\sigma_N}\ket{\sigma_1}\otimes\ket{\sigma_2}\otimes\ldots\otimes\ket{\sigma_N} \ ,
\end{equation}
where $\left\{M^{\sigma_i}\right\}$ is a set of matrices (the MPS matrices) characterising the $i$th site, see Ref.~\cite{schollwock_density-matrix_2011} for a detailed discussion. The calculation of $\rho_1$ in MPS formalism is particularly efficient if the chain is of the so-called ``orthonormal'' form~\cite{schollwock_density-matrix_2011}. Under this condition, the MPS sets of matrices $i=2,\ldots,N$ satisfy the orthogonality properties $\sum_{\sigma_i}\left(M^{\sigma_i}\right)^\dag M^{\sigma_i} = I_i$, and the calculation of $\rho_1$ reduces to
\begin{equation}
	\rho_{1} = \sum_{\sigma_1,\sigma_1'}M^{\sigma_1}\left(M^{\sigma'_1}\right)^\dag\ket{\sigma_1}\bra{\sigma_1'} \ .
\end{equation}
Analogously, the density operator of the $i$th site of the chain reduces to
\begin{equation} 
	\label{Eq: App App reduced density matrix for a single site}
	\rho_{i} = \sum_{\sigma_i\sigma_i'}M^{\sigma_i}\left(M^{\sigma'_i}\right)^\dag\ket{\sigma_i}\bra{\sigma_i'} \ ,
\end{equation}
from which the square of the density operator reads
\begin{equation}
	\rho_i^2 = \sum_{\sigma_i,\sigma_i',\sigma_i''}M^{\sigma_i}\left(M^{\sigma'_i}\right)^\dag M^{\sigma_i'}\left(M^{\sigma''_i}\right)^\dag\ket{\sigma_i}\bra{\sigma_i''} \ .
\end{equation}
The purity $\mathcal{P}_i$ of the $i$th chain site in MPS formalism therefore reads
\begin{equation}
	\mathcal{P}_i = \Tr(\rho_i^2) = \Tr\left\{\sum_{\sigma_i,\sigma_i'}M^{\sigma_i}\left(M^{\sigma'_i}\right)^\dag \left[M^{\sigma_i}\left(M^{\sigma'_i}\right)^\dag \right]^\dag\right\} \ .
\end{equation}

\end{document}